\documentclass[preprint,12pt]{elsarticle}

\usepackage{lineno,hyperref}
\modulolinenumbers[5]

\usepackage{amsmath}
\usepackage{amssymb}

\journal{}

\biboptions{numbers,sort&compress}

\begin{document}

\begin{frontmatter}

\title{Possible influence of the Kuramoto length in a photo-catalytic water splitting reaction revealed  
by Poisson--Nernst--Planck equations 
involving ionization in a weak electrolyte}

\author{Yohichi Suzuki}
\author{Kazuhiko Seki\corref{mycorrespondingauthor}}
\address{Nanomaterials Research Institute(NMRI), 
National Institute of Advanced Industrial Science and Technology (AIST)\\
AIST Tsukuba Central 5, Higashi 1-1-1, Tsukuba, Ibaraki, Japan, 305-8565}


\cortext[mycorrespondingauthor]{Corresponding author}
\ead{k-seki@aist.go.jp}


\begin{abstract}
We studied ion concentration profiles and the charge density gradient caused by electrode reactions in weak electrolytes by using the Poisson--Nernst--Planck equations
without assuming charge neutrality. 
In weak electrolytes, only a small fraction of molecules is ionized in 
bulk. 
Ion concentration profiles depend on not only ion transport but also  
the ionization of molecules. 
We considered the ionization of molecules and ion association in weak electrolytes and 
obtained analytical expressions for ion densities, electrostatic potential profiles, and ion currents. 
We found the case that the total ion density gradient 
was given by the Kuramoto length 
which characterized the distance over which an ion diffuses before association. 
The charge density gradient is characterized by the Debye length 
for 1:1 weak electrolytes. 
We discuss the role of these length scales for efficient water splitting reactions using 
photo-electrocatalytic electrodes. 
\end{abstract}


\end{frontmatter}

\section{Introduction}
\label{sec:I}
The Poisson-Nernst-Planck (PNP) equations have been used to describe a wide range of transport phenomena 
from electrons and holes in semiconductors to ions in electrolytes. \cite{Levich_62,Newman_04,Bard_book_01,Mafe_86,Bier_83,BUCK_84,Cohen_65,Rubinstein_79,Kim_10} 
The conservation of charge for each ion species and electrostatic interactions among charge carriers are 
treated by the PNP equations in a self-consistent manner. 
The PNP equations take into account the drift currents   
due to the electric fields generated by the distribution of charge carriers. 
The effect is substantial {\it e.g.} when electrode reactions generate charge density gradients.  
The PNP equations can be applied to obtain 
concentration profiles and an electro-static potential generated by electrode reactions. 
Although the PNP equations are used for the study of coupled effects between electric fields and charge carrier transports, 
they are nonlinear and solved mainly by numerical methods. 
However, for certain cases, approximate analytical solutions have also been obtained for strong electrolytes. \cite{Newman_04,Henderson_07,Bard_book_01,Dickinson2011,Jackson,Bazant_04,Rubinstein_79,Kim_10}
Using the PNP equations, it has been shown that 
the spatial dimensions of concentration gradient can be many orders of magnitude larger than 
the characteristic length scale of charge density profiles given by the Debye length in strong electrolytes. \cite{Levich_62,Newman_04,Mafe_86,Bard_book_01,Bazant_04}

In weak electrolytes, only a small fraction of molecules are ionized in 
bulk. 
Ion concentration profiles depend on ion transport and  
the ionization of molecules. 
It should be noted that depleted ions 
in the vicinity of the electrode can be replenished by the ionization of molecules and diffusion from the bulk 
in weak electrolytes. 
The current at the interface between the electrolyte and electrode 
can be much larger than that in bulk due to the molecule ionization.  

The Nernst-Planck equation is the continuity equation representing the 
conservation of charge for each ion species, 
where dissociation and association of ions are not considered. 
For weak electrolytes, the PNP equations have been extended to describe ionization and recombination of ions. \cite{Bier_83,Grossman_76,Nikonenko_10,Cheng_11,Zabolotskii_12,KHARKATS_91,Nielsen_14,Jin_14,Xiang_16,Femmer_15,VANPARYS_10,Kodym_16,AOKI_13}
In the extended PNP (e-PNP) equations, ion concentrations become non-conservative by 
local dissociation and association of ions in bulk phase.  

In this paper, we study charge transport induced by electrode reactions in weak electrolytes using the e-PNP equations without assuming a priori charge neutrality.    
Photo-electrochemical (PEC) conversion of water can be an example, but 
the fundamental results apply to other electrode reactions. 
We show that the gradient of total ion density (the sum of cation and anion concentrations) 
is characterized by either 
the Kuramoto length \cite{Kuramoto_73,Kuramoto_74,Nitzan_74,Nicolis_84,Kitahara_90,WAKOU_02} 
or the Debye length depending on the situation, while the gradient of charge density (the difference between cation and anion concentrations) 
is characterized by 
the Debye length in binary monovalent weak electrolytes. 
The Kuramoto length characterizes the length scale of 
local density fluctuations around a uniform concentration state. \cite{Kuramoto_73,Kuramoto_74,Nitzan_74,Nicolis_84,Kitahara_90,WAKOU_02} 
In weak electrolytes,  
the Kuramoto length is given by the length scale of diffusive migration of ions within its life-time; 
the life-time is determined by the association rate of ions. \cite{Kuramoto_73,Kuramoto_74,Nitzan_74,Nicolis_84,Kitahara_90,WAKOU_02} 
Our results indicate that the ion density gradients can also be characterized by the Kuramoto length 
when both cations and anions are discharged at the electrode in binary monovalent weak electrolytes.
For this case, ion density drop caused by electrode reactions is recovered by ion density fluctuations 
localized within the Kuramoto length.  
We discuss the efficiency of water splitting reactions using 
photo-electrocatalytic electrodes in terms of 
the Kuramoto length and the Debye length.  
We also discuss the overpotential related to a charge density gradient near the electrode and 
show that it can be reduced when both cations and anions are discharged at the electrode. 

%\newpage
%%%%%%%%%%%%%%%%%%%%%%%%%%%%%%%%%%%%%%%%%%%%%%%%%%%%%%%%%%%%%%%%%%%%%%
% Theory %%%%%%%%%%%%%%%%%%%%%%%%%%%%%%%%%%%%%%%%%%%%%%%%%%%%%%%%%%%%%%
%%%%%%%%%%%%%%%%%%%%%%%%%%%%%%%%%%%%%%%%%%%%%%%%%%%%%%%%%%%%%%%%%%%%%%
%\setcounter{equation}{0}
\section{Theory}
\label{sec:II}
%%%%%%%%%%%%%%%%%%%%%%%%%%%%%%%%%%%%%%%
\begin{figure}
\centerline{
\includegraphics[width=1\columnwidth]{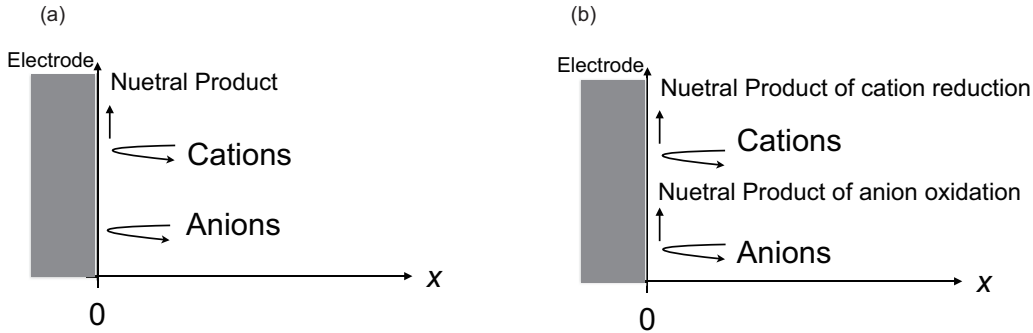}
}
\caption{Sketch of the model of ion discharge reactions at the electrode surface. 
The distance from the electrode is denoted by $x$. 
(a) Cations are reduced at the electrode surface, and the neutral products are dissolved into the solution.  
Anions are reflected at the electrode surface.
(b) Cations are reduced, and anions are oxidized at the electrode surface. 
The both neutral products are removed from the electrode surface. 
For photo-electrocatalytic (PEC) water splitting using a single photocatalyst, 
cations and the neutral products correspond to 
H$_3$O$^+$ and H$_2$, respectively;  
anions and the neutral products correspond to 
OH$^-$ and O$_2$, respectively. 
}
\label{fig1}
\end{figure}
%%%%%%%%%%%%%%%%%%%%%%%%%%%%%%%%%%%%%%%
As shown in Fig. \ref{fig1},  
we considered the thermal motion of cation and anion under Coulombic interactions. 
The one-dimensional (1D) $x$-coordinate was introduced by assuming a uniform distribution of ions 
in the plane parallel to the electrode surface. 
The origin of the $x$-coordinate was the electrode surface, 
and the $x$-coordinate was normal to the electrode surface. 
The concentration and diffusion constant of cations are denoted as $n_+$ and $D_+$, respectively. 
The concentration and diffusion constant of anions are denoted as $n_-$ and $D_-$, respectively, while the electric field is denoted as $E$. 
Each species moves by electromigration and diffusion. 
For simplicity, we consider a binary monovalent electrolyte (1:1 electrolyte). 
The concentration and the diffusion constant of the undissociated neutral compound are denoted as $m$ and $D_m$, respectively.
At a sufficiently large distance $L$ away from the electrode, 
the electrolyte is neutral. 
The external electric field is not applied. 
We consider ion discharge reaction at the electrode followed by product formation. 
Products of electrode reactions are assumed to be removed. 
An example is hydrogen evolution reaction by photo-electrocatalytic (PEC) water splitting. \cite{Wang_16,Wang_17}
For PEC water splitting, cations, anions, and undissociated molecules correspond to 
H$_3$O$^+$, OH$^-$, and water molecules, respectively. 
We mainly consider the case that one of the ion species is discharged at the electrode and 
the other species are inert and reflected at the electrode as described in Fig. \ref{fig1} (a).
In Sec. \ref{sec:BC}, we consider the special case 
that both cation and anion species are discharged at the electrode 
as described in Fig. \ref{fig1} (b). 
The situation corresponds to 
overall water splitting into H$_2$ and O$_2$ using a single photocatalyst such as 
GaN-ZnO and ZnGeN$_2$-ZnO under light illumination. \cite{Lee_07,Maeda_06}
For simplicity, we assume that reactive sites on the electrode are structureless.  
In experiments, the electrode can be regarded as homogeneous when the electrode is modified with
molecular or metal complex cocatalyst to 
realize both oxidation and reduction reactions on the electrode.
Some electrodes with nano-particulate photocatalysts could also be regarded as homogeneous. 

Ion transport was described by the Nernst--Planck equations 
and the electric field satisfied the Poisson equation. 
By taking into account the association and dissociation reactions of ions
with diffusion and ion migration under the electric field $E(x)$, 
the governing equations in the bulk phase are given by 
\cite{Bier_83,Grossman_76,Nikonenko_10,Cheng_11,Zabolotskii_12,KHARKATS_91,Nielsen_14,Jin_14,Xiang_16,VANPARYS_10,Kodym_16,AOKI_13}
\begin{align}
\frac{\partial}{\partial t} n_+&=
 \frac{\partial}{\partial x} D_+ \left( 
\frac{\partial n_+}{\partial x}  -  
\frac{eEn_+}{k_{\rm B} T} 
\right) +k_d m - k_a n_+ n_-, 
\label{eq:NPR1} \\
\frac{\partial}{\partial t} n_-&=
 \frac{\partial}{\partial x} D_- \left( 
\frac{\partial n_-}{\partial x}  + 
\frac{eEn_-}{k_{\rm B} T} 
\right) +k_d m - k_a n_+ n_- , 
\label{eq:NPR2} \\
\frac{\partial}{\partial t} m&=
 D_m 
 \frac{\partial^2}{\partial x^2}m -k_d m + k_a n_+ n_-, 
\label{eq:NPR3} \\
\frac{\partial}{\partial x} E(x) &= \frac{4\pi e}{\epsilon} \left(n_+ - n_-\right),
\label{eq:C}
\end{align}
where $k_d$ and $k_a$ denote the dissociation rate constant and the association rate constant, respectively. 
Equation (\ref{eq:C}) is the Poisson equation. 
$-D_+  
(\partial n_+)/(\partial x)$ 
and $D_+ eEn_+/(k_{\rm B} T)$ on the right-hand side of Eq. (\ref{eq:NPR1}) represent the diffusive flux 
and the ion electro-migration by the electric field $E$, respectively.
$D_+ /(k_{\rm B} T)$ can be regarded as the mobility using 
the Einstein relation. 
When the electrostatic potential is controllable by potentiostat operation, 
Eqs. (\ref{eq:NPR1})-(\ref{eq:C}) may be more conveniently expressed by using the electrostatic potential 
rather than using the electric field. 
However, we study the ion currents and the electrostatic potential induced by 
the ion discharge reactions at the electrode without imposing the external electric field. 
In this case, the electrostatic potential is not a controllable parameter. 
To avoid a redundant integration constant associated with the electrostatic potential we express 
Eqs. (\ref{eq:NPR1})-(\ref{eq:C}) using the electric field. 

In the following, we consider steady states. 
The left-hand sides of Eqs. (\ref{eq:NPR1}) and (\ref{eq:NPR2}) are zero in steady states.  
The electrostatic potential difference relative to the electrostatic potential at $L$ can be defined by
\begin{align}
V(x)= - \int_{L}^{x} dx_1 E(x_1),  
\label{eq:V}
\end{align}
and can be calculated by solving Eqs. (\ref{eq:NPR1})--(\ref{eq:C}) under proper boundary conditions. 
The total potential difference between the electrode surface and bulk (overpotential) is $V(0)$.

\subsection{Boundary Conditions}
\label{sec:II_A}

 At a sufficiently large distance away from the electrode surface, the ion concentrations are not affected by 
surface reactions. 
 Correspondingly, 
we set boundary conditions at $x=L$ and assumed equal concentrations of cations and anions given by $n_b$. 
Boundary conditions at $x=L$ are given by
\begin{align}
\left. n_+\right|_{x=L}=n_b, 
\label{eq:BCpp1}\\
\left. n_-\right|_{x=L}=n_b, 
\label{eq:BCpm1}\\
\left. m \right|_{x=L}=m_b, 
\label{eq:BCpml1}
\end{align}
where $k_d m_b=k_a n_b^2$ is satisfied in the bulk. 
We consider the situation that the external electric field is absent in the bulk  
\begin{align}
E(L)=0. 
\label{eq:BCEM}
\end{align}
In ideal situation of photo-electrochemical (PEC) water splitting, 
the external electric field is not applied. 
We also assumed a metal electrode, where 
the net image charge in the electrode is the same as the net charge in the electrolyte system. 
Therefore, the electric fields at the boundaries 
enclosing both the real charges in the electrolyte and the image charges are zero. 
This situation can also be viewed from another perspective.  
Because we consider neutral products from a neutral electrolyte by electrode reactions, 
the net charge  in the system including those induced in the electrode should be neutral. 
In this case, electric fields at some distance sufficiently far from the electrodes should be zero.\cite{BUCK_84}

We consider the case where cations are reduced at the interface between the electrolyte and electrode and the 
anions are reflected as shown in Fig. \ref{fig1} (a). 
Boundary conditions at the cathode located at $x=0$ can be expressed as
\begin{align}
n_+ (0)&=n_s, 
\label{eq:BCmp1}\\
\left. 
D_-\left(\frac{\partial n_-}{\partial x}  +
\frac{eE n_-}{k_{\rm B} T}
\right)
\right|_{x=0}&=0,  
\label{eq:BCmm2}\\
\left. 
\frac{\partial m}{\partial x}  
\right|_{x=0}&=0. 
\label{eq:BCmml1}
\end{align}
Equation (\ref{eq:BCmml1}) indicates that undissociated molecules are reflected by the cathode. 
The left-hand side of (\ref{eq:BCmm2}) represents the concentration current of anions. 
Equation (\ref{eq:BCmm2}) indicates that anions are reflected by the cathode. 
In Eq. (\ref{eq:BCmp1}),  
$n_s$ represents the net effect of surface reactions on $n_+(0)$.  
As a plausible boundary condition for irreversible surface reactions, we considered that $n_s=0$ on the right-hand side of 
Eq. (\ref{eq:BCmp1}). 
The boundary condition is suitable to study the case when the overall reactions are limited 
by ion transport to the electrode surface. 
In general, we set $n_s$ being constant on the right-hand side of 
Eq. (\ref{eq:BCmp1}). 
%%% up to this point

%%%%%%%%%%%%%%%%%%%%%%%%%%%%%%%%%%%%
\subsection{Dimensionless Equations}
\label{sec:II_B}

We can express concentrations of cations and anions in dimensionless units as
\begin{align}
N_+=\frac{n_+}{n_b}, N_-=\frac{n_-}{n_b}, N_s=\frac{n_s}{n_b} \mbox{ and } N_m=\frac{m}{m_b}. 
\label{eq:DLN}
\end{align}
For convenience, we introduce \cite{Jackson}
\begin{align}
\bar{D} &= \frac{2D_+ D_-}{D_+ + D_-}, \Delta = \frac{D_+ - D_-}{D_+ + D_-}. 
\label{eq:DLD}
\end{align}
Using the above definitions, the original diffusion constants can be obtained by
$
D_+ = \bar{D}/ \left(1- \Delta \right)$, 
and 
$D_- = \bar{D}/ \left(1+ \Delta \right)$. 
By using  
the Debye length given by \cite{Debye_23,Jackson,Levin_02,Bazant_04,Andelman} 
\begin{align}
\lambda_D=\left( \frac{\epsilon k_{\rm B} T}{8 \pi e^2 n_b}
\right)^{1/2},
\label{eq:Debye}
\end{align}
we can express the characteristic electric field strength and diffusion time as \cite{Jackson}
\begin{align}
E_0=\frac{k_{\rm B} T}{e \lambda_D}=\left( \frac{8 \pi n_b k_{\rm B} T}{\epsilon}
\right)^{1/2},
t_0=\frac{\lambda_D^2}{\bar{D}} =\frac{\epsilon k_{\rm B} T}{8 \pi e^2 n_b \bar{D}}.
\label{eq:iDL}
\end{align}
We also introduce dimensionless variables given by
\begin{align}
y=\frac{x}{\lambda_D}, \tau=\frac{t}{t_0}, \xi=\frac{E}{E_0}. 
\label{eq:DLx}
\end{align}
Finally, we express the reaction rate constants in the dimensionless form, 
\begin{align}
\bar{k}_a=\frac{\lambda_D^2 k_a n_b}{\bar{D}} ,
\bar{k}_d=\frac{\lambda_D^2 k_d}{\bar{D}} .
\label{eq:rc}
\end{align}
Here, $\bar{k}_a$ and $\bar{k}_d$ represent the dimensionless ion association rate constant and 
ion dissociation rate constant, respectively. 
Equation (\ref{eq:NPR1})--(\ref{eq:C}) can be rewritten using the dimensionless variables as
\begin{align}
\frac{\partial^2 N_+}{\partial y^2} - 
\frac{\partial}{\partial y} \left( \xi N_+\right) &=  \left(1-\Delta\right) \left[ \frac{\partial N_+}{\partial \tau} 
-  
\bar{k}_a\left( N_m - N_+ N_- \right) \right],
\label{eq:DLNp1}\\
\frac{\partial^2 N_-}{\partial y^2} + 
\frac{\partial}{\partial y} \left( \xi N_-\right) &= 
 \left(1+\Delta\right)  \left[ \frac{\partial N_-}{\partial \tau}
- 
\bar{k}_a\left( N_m - N_+ N_- \right) \right],
\label{eq:DLNm1}\\
\frac{\partial^2 N_m}{\partial y^2}  &= 
\frac{\bar{D}}{D_m} \left[ \frac{\partial N_-}{\partial \tau}
+ 
\bar{k}_d \left( N_m - N_+ N_- \right) \right],
\label{eq:DLM1}\\
\frac{\partial}{\partial y} \xi &= \frac{1}{2} \left( N_+ - N_- \right). 
\label{eq:DLNE1}
\end{align}
Similarly, boundary conditions at $y=L/\lambda_D$ are given by
\begin{align}
\left. N_+\right|_{y=L/\lambda_D}=1, 
\label{eq:BCpp2}\\
\left. N_-\right|_{y=L/\lambda_D}=1,
\label{eq:BCpm2}\\
\left. N_m\right|_{y=L/\lambda_D}=1,
\label{eq:BCpml2}
\end{align}
and boundary conditions at $y=0$ [Eqs. (\ref{eq:BCmp1})-(\ref{eq:BCmml1})] are given by
\begin{align}
N_+(0) &=N_s, 
\label{eq:BCm1}\\
\left. 
\left(\frac{\partial N_-}{\partial y}  +
\xi N_-
\right)
\right|_{y=0}&=0 , 
\label{eq:BCm2}\\
\left. 
\frac{\partial N_m}{\partial y}  \right|_{y=0}&=0 . 
\label{eq:BCml2}
\end{align}
Finally, the boundary condition for the electric field given by Eq. (\ref{eq:BCEM}) 
can be expressed as 
\begin{align}
\left. \xi \right|_{y=L/\lambda_D}=0. 
\label{eq:BCC_xi}
\end{align}
The potential difference given by Eq. (\ref{eq:V}) can be expressed as 
\begin{align}
V(y)= - E_0 \lambda_D \int_{y=L/\lambda_D}^{y} dy \xi(y). 
\label{eq:V1_0o}
\end{align}

Below we consider the case where the condition
\begin{align}
\bar{k}_a \frac{n_b}{m_b} \frac{\bar{D}}{D_m} \ll 1
\label{eq:wel}
\end{align}
is satisfied;  
 under the condition and using $k_d m_b=k_a n_b^2$, 
 it can be shown that 
the right-hand side of Eq. (\ref{eq:DLM1}) can be ignored in steady states. 
For this case,    
the dimensionless concentration of undissociated molecules can be approximated by $N_m=1$ 
from the solution of diffusion equation under the boundary conditions of Eqs. (\ref{eq:BCpml2}) and (\ref{eq:BCml2}). 
This situation represents that 
the ratio of dissociated to undissociated neutral compound given by 
$n_b/m_b$ is so small in weak electrolytes that 
the density of the neutral compound can be approximated as constant. 
Obviously, Eq. (\ref{eq:wel}) is satisfied for water. 
For general weak electrolytes, additional condition for the concentration of the undissociated neutral compound 
may be needed. 
By using the degree of ionization $\alpha$, we have $n_b/m_b=\alpha/(1-\alpha)$, 
where the numerator and the denominator indicate the increase of the amount of cation 
and the decrease in the amount of undissociated molecules for the extent of ionization reaction, respectively. 
When we consider the dissociation of an initial amount $n_0$ of the undissociated compound, 
the equilibrium dissociation constant is expressed as $K_a=n_0 \alpha^2/(1-\alpha)$, 
where $n_0$ is divided by the standard concentration 1 mol/L. 
In weak electrolytes,  $n_0$ typically satisfies the relation $n_0 \gg K_a/4$ and 
we obtain $\alpha \approx \sqrt{K_a/n_0}$. 
By substituting $n_b/m_b\approx \alpha \approx \sqrt{K_a/n_0}$, 
Eq. (\ref{eq:wel}) can be rewritten as 
$K_a/n_0  \ll (D_m/\bar{D})^2/\bar{k}_a^2$. 
Using $D_m/\bar{D}\sim 10$ as a typical values of the diffusion constants of Carboxylic acids \cite{Albery_67} and
$1<\bar{k}_a<10$, we find $K_a/n_0\ll 10^{-2}\sim 10^{-4}$. 
For most weak electrolytes satisfying $K_a\leq 10^{-4}$, the condition can be fulfilled under normal values of $n_0$. 
Below, we assume that the condition is satisfied and put $N_m=1$ in 
Eqs. (\ref{eq:DLNp1})-(\ref{eq:DLNm1}).

%%%%%%%%%%%%%%%%%%%%%%%%%%%%%%%%%%%%%%%%%%%%%
\section{Analytical Results}
%%%%%%%%%%%%%%%%%%%%%%%%%%%%%%%%%%%%%%%%%%%%%
\label{sec:III}
When ion discharge reactions proceed at the electrode-electrolyte interface 
in a steady state, 
the time-derivatives in Eqs. (\ref{eq:DLNp1})-(\ref{eq:DLNm1}) can be set zero. 
Here, we derive time independent approximate solutions of 
Eqs. (\ref{eq:DLNp1})-(\ref{eq:BCC_xi}) by assuming $N_m=1$ as explained in the previous section. 
As explained later, the steady--state analytical solution is obtained under a certain condition relevant to weak electrolytes. 
A separate consideration is needed for strong electrolytes as shown in Appendix A. 

Because we are interested in the deviations from equilibrium concentrations induced by surface reactions, 
we introduce 
sum and change as \cite{Jackson}
$
S= N_+ + N_-, C= N_+ - N_- $, where 
$S$ and $C$ characterize the total mass density and charge density, respectively.  
To investigate the deviations from equilibrium states, we introduce 
\begin{align}
\delta S = S-2, \delta C=C.  
\label{eq:devSC_0}
\end{align}
In steady states, Eqs. (\ref{eq:DLNp1}), (\ref{eq:DLNm1}) and 
(\ref{eq:DLNE1}) can be rewritten using these variables as 
\begin{align}
\frac{\partial^2 \delta S}{\partial y^2} - 
\frac{\partial}{\partial y} \left( \xi \delta C\right) &= 
 2 \bar{k}_a\left[\delta S +\frac{1}{4} \left(\delta S^2-\delta C^2\right) \right]
,
\label{eq:sum1_0}\\
\frac{\partial^2 \delta C}{\partial y^2} - 
\frac{\partial}{\partial y} \left[ \xi (\delta S+2) \right] &= 
-2 \Delta \bar{k}_a \left[\delta S +\frac{1}{4} \left(\delta S^2-\delta C^2\right) \right]
,
\label{eq:change1_0}\\
\frac{\partial}{\partial y} \xi &= \frac{1}{2} \delta C , 
\label{eq:xiC1_0}
\end{align}
where $N_m=1$ is set by assuming the condition 
Eq. (\ref{eq:wel}) is satisfied. 
In a weak electrolyte,  the dissociation of a small amount of the solute can be represented by 
the right-hand sides of Eqs. (\ref{eq:sum1_0}) and (\ref{eq:change1_0}).

The boundary conditions at $y=L/\lambda_D$ can be written as
\begin{align}
\left. \delta S\right|_{y=L/\lambda_D}=0, 
\label{eq:BCS1_0o}\\
\left. \delta C\right|_{y=L/\lambda_D}=0. 
\label{eq:BCC2_0o}
\end{align}
The boundary conditions at $y=0$ can be expressed as 
\begin{align}
\left. \frac{\delta S+\delta C}{2}\right|_{y=0}&= N_s -1, 
\label{eq:BCSm1_0o}\\
\left. \frac{\partial }{\partial y} (\delta S-\delta C)+\xi(2+\delta S-\delta C)\right|_{y=0}&=0.  
\label{eq:BCSm3_0o}
\end{align}
The boundary condition for the electric field is given by Eq. (\ref{eq:BCC_xi}).

By linearizing the above equations, we solved the following set of equations, 
\begin{align}
\frac{\partial^2 \delta S}{\partial y^2}  &= 
 2 \bar{k}_a \delta S 
,
\label{eq:add11}\\
\frac{\partial^2 \delta C}{\partial y^2} - 
2\frac{\partial}{\partial y} \xi  &= 
-2 \Delta \bar{k}_a  \delta S 
. 
\label{eq:add12o}
\end{align}
By substituting Eq. (\ref{eq:xiC1_0}) into Eq. (\ref{eq:add12o}), we obtain 
\begin{align}
\frac{\partial^2 \delta C}{\partial y^2} -\delta C= -2 \Delta \bar{k}_a  \delta S 
. 
\label{eq:add12}
\end{align}
If the right-hand side is zero, Eq. (\ref{eq:add12}) 
is equal to the equation derived by Debye and Falkenhagen in the steady state. \cite{Debye_28}
The right-hand side of Eq. (\ref{eq:add12}) 
reflects ionization of molecules. 
When the ion concentrations are conserved, the Debye--Falkenhagen equation
is obtained, and the length scale of the charge density variation is characterized by the Debye length. 
Equation (\ref{eq:add12}) indicates that 
the charge density variation is still characterized by the Debye length 
even when the ion concentrations are non-conserved by ionization of molecules:  
the ionization effect is represented by the inhomogeneous term in Eq. (\ref{eq:add12}) 
when the linearization approximation is used. 
The validity of the linearization will be discussed later. 
Hereafter, we solve Eqs. (\ref{eq:add11}), (\ref{eq:add12}) and (\ref{eq:xiC1_0}) using the boundary conditions given by 
Eqs. (\ref{eq:BCS1_0o})-(\ref{eq:BCSm3_0o}) and (\ref{eq:BCC_xi}). 

The solution of Eq. (\ref{eq:add11}) can be expressed as 
\begin{align}
\delta S(y)=C_s \exp \left(-\sqrt{2 \bar{k}_a}\, y\right)
\label{eq:deltaSsol}
\end{align}
using the boundary condition given by Eq. (\ref{eq:BCS1_0o}) in the limit of 
$L \rightarrow \infty$, where $C_s$ is an integration constant.  
Using the boundary condition given by Eq. (\ref{eq:BCC2_0o}) in the limit of 
$L \rightarrow \infty$, we obtained from Eq. (\ref{eq:add12}) 
\begin{align}
\delta C (y)= - 2\bar{k}_a \Delta /(2\bar{k}_a-1) C_s \exp \left(-\sqrt{2 \bar{k}_a}y\right) +C_c \exp \left(-y\right) ,
\label{eq:deltaCsol}
\end{align}
where $C_c$ is an integration constant. 
By substituting them into Eqs. (\ref{eq:BCSm1_0o}) and (\ref{eq:BCSm3_0o}) and linearizing the equations, 
we obtain $C_s=0$ and $C_c=-2(1-N_s)$ up to the linear order in $1-N_s$. 
By substituting $\delta C$ and integrating Eq. (\ref{eq:xiC1_0}), we found that
$\xi(y)=(1-N_s) \exp \left(-y\right) + C_\xi$,  
where $C_\xi$ is an integration constant.
The boundary condition given by Eq. (\ref{eq:BCC_xi}) can be expressed as 
$\xi(y) \rightarrow 0$ in the limit of $L \rightarrow \infty$. 
Therefore, we find that $C_\xi=0$.
These results can be written as
\begin{align}
\delta C(y) &= -2 (1-N_s) \exp (-y), 
\label{eq:21}\\
\xi(y)&= (1-N_s) \exp (-y). 
\label{eq:22}
\end{align}
The electric field strength can be written as 
\begin{align}
E(y)=\frac{k_{\rm B} T}{e \lambda_D}  (1-N_s) \exp (-y).
\label{eq:22_1}
\end{align}
From Eq. (\ref{eq:V1_0o}) and taking the limit of $L \rightarrow \infty$, the potential profile can be obtained as 
\begin{align}
V(y)
&= \frac{k_{\rm B} T}{e} \left(1-N_s\right) \exp \left(-y\right).
\label{eq:23}
\end{align}
Using $N_+=(2+ \delta C)/2$ and $N_-=(2- \delta C)/2$, we obtained the ionic concentration profiles as 
\begin{align}
N_+ (y)&= 1-\left(1-N_s\right) \exp \left(-y\right), 
\label{eq:24}\\
N_- (y)&= 1+\left(1-N_s\right) \exp \left(-y\right).  
\label{eq:24_1}
\end{align}
Exponential screening of Eqs. (\ref{eq:24})-(\ref{eq:24_1}) reminds us of 
the exponential decrease of electric field formed near 
charged surfaces in strong electrolytes under local equilibrium \cite{Debye_23,Levin_02,Bazant_04,Andelman,Barbero_02};  
the solution for a certain boundary condition is known as the Debye--H\"{u}ckel theory.
However, the situation considered here is different from that of the Debye--H\"{u}ckel theory.
The Debye--H\"{u}ckel theory is based on the Poisson--Boltzmann equation, where 
ion concentrations are assumed to obey local equilibrium. 
Here, we solved the e-PNP equations where the effect of ion currents is taken into account and 
deviations from the local equilibrium distributions can be studied. 

The dimensionless concentration current density in Eq. (\ref{eq:BCm1}) was obtained from Eqs. (\ref{eq:24}) and (\ref{eq:22}) as 
\begin{align}
\left(\frac{\partial N_+}{\partial y}  -
\xi N_+
\right) =\left(1-N_s \right)^2 \exp (-2 y). 
\label{eq:current_liear_non}
\end{align}
The ion current density can be expressed as 
\begin{align}
e \frac{D_+}{\lambda_D} n_b \left(1-\frac{n_s}{n_b} \right)^2 \exp (-2 x/\lambda_D), 
\label{eq:current_liear}
\end{align}
and its value at the electrode surface is given by 
\begin{align}
e \frac{D_+}{\lambda_D} n_b \left(1-\frac{n_s}{n_b} \right)^2 ,
\label{eq:current_liear_0}
\end{align}
using a linearized approximation.
 
The above results were derived by linearizing the Poisson--Nernst--Planck equations with the association and dissociation reactions. 
The linearization is valid if $|\partial \left( \xi \delta C\right)/\partial y |$ on the left-hand side  of Eq. (\ref{eq:sum1_0}) is smaller than 
$2 \bar{k}_a |\delta S|$ on the right-hand side. 
The linearized solutions indicate $\delta C\partial  \xi /\partial y = \xi \partial  \delta C /\partial y$ and 
$|\partial \left( \xi \delta C\right)/\partial y | = 2\delta C\partial \xi /\partial y$. 
The magnitude of charge density $\delta C$ would decrease by increasing the distance from the electrode surface and 
$\delta C\partial  \xi /\partial y = \delta C^2/2 \approx 2\left(1-N_s\right)^2$ near the electrode surface. 
If the electrode is perfectly reactive, we have 
$\delta C\partial  \xi /\partial y \approx 2$ and 
$|\partial \left( \xi \delta C\right)/\partial y | = 2\delta C\partial  \xi /\partial y \approx 4$ 
near the electrode surface. 
By considering $\delta S \approx N_- \approx |\delta C|$, 
we also obtain  
$\delta S \approx 2$ near the electrode surface if the electrode is perfectly reactive.  
The condition for linearization given by $|\partial \left( \xi \delta C\right)/\partial y | \ll 2 \bar{k}_a |\delta S|$ can be
expressed as $\bar{k}_a >1  $. 
Otherwise, the coupling between $\xi$ and $\delta C$ in 
Eq. (\ref{eq:sum1_0}) cannot be ignored. 
The linearization condition will be numerically studied in Sec. \ref{sec:V}.

%%%%%%%%%%%%%%%%%%%%%%%%%%%%%%%%%%%%%%
\section{Interpretation of the linearization condition}
\label{sec:IV}

As mentioned above, 
the linearization condition of the e-PNP equations is given by  $\bar{k}_a > 1$. 
The linearization condition can be interpreted in terms of two time scales 
using the definition $\bar{k}_a=\lambda_D^2 k_a n_b/\bar{D}$. 
The value of $1/(k_a n_b)$ gives the time scale of recovering charge neutrality by bulk reactions. 
The time scale of diffusion over the Debye length is given by  
$t_0=\lambda_D^2/\bar{D}$; the Debye length characterizes the spatial extent of the deviation from charge neutrality.  
$\bar{k}_a > 1$ indicates that 
the time scale of bulk reactions is shorter than that of diffusion over the Debye length.  
Therefore, charge neutrality is mainly recovered by ion association and dissociation reactions when $\bar{k}_a>1$. 
In the opposite limit of  
$\bar{k}_a<1$, 
charge neutrality is mainly recovered by ion diffusion. 

As shown below, the linearization condition is satisfied when ion association is diffusion limited. 
The diffusion-limited ion association rate can be relevant for weak electrolytes, 
where the intrinsic association rate constants are large, and the dissociation rate constants are small. 
For example, the association rate of water can be estimated 
as $k_a=1.3 \times 10^{11} \mbox{L/(mol s)}$ using the diffusion limited bulk rate constant 
$4 \pi (D_+ + D_-)R_{\rm eff}$ together with the effective reaction radius given by  $R_{\rm eff}= |r_c|/[1- \exp(-|r_c|/R)]$, 
\cite{Debye_42,Eigen_58,Stillinger_78,Natzle_85} 
where $|r_c|=e^2/(\epsilon k_{\rm B} T)$ is the Onsager radius and 
the diffusion constants of H$_3$O$^+$ and OH$^-$ are 
$9.4 \times 10^{-9}$ m$^2$/s and $5.3 \times 10^{-9}$ m$^2$/s, respectively. \cite{Silbey}
In the above estimation, we also used $\epsilon=80$ and 
$R=0.75$ nm. \cite{Eigen_58,Stillinger_78}
The estimated $k_a$ value is close to the literature value given by  
$k_a=1.4 \times 10^{11} \mbox{L/(mol s)}$. \cite{Eigen_58,Stillinger_78,McQuarrie}
This indicates that the effect of diffusive interaction can be ignored owing to 
the low degree of ionization in weak electrolytes. \cite{Traytak_07}
The similar estimation is also possible for other weak electrolytes. \cite{Brown_17} 
By noticing that the Debye length can be expressed using the Onsager length as  
\begin{align}
\lambda_D= \left( 
8\pi |r_c| n_b
\right)^{-1/2}, 
\label{eq:Debye_Onsager}
\end{align} 
$\bar{k}_a=\lambda_D^2 k_a n_b/\bar{D}$ can be expressed as 
\begin{align}
\bar{k}_a= \frac{1}{4} \frac{(D_+ + D_-)^2}{D_+ D_-}
\frac{1}{1- \exp(-|r_c|/R)}. 
\label{eq:barka_1}
\end{align}
Because the minimum value of 
$(D_+ + D_-)^2/(D_+ D_-)$ 
is obtained when $D_+=D_-$, 
we find 
\begin{align}
\bar{k}_a>
\frac{1}{1- \exp(-|r_c|/R)}>1  
\label{eq:barka_2}
\end{align}
for given values of $\epsilon$ and the reaction radius. 
For water, we find $\bar{k}_a=2.0$ and the linearization condition $\bar{k}_a>1$ is satisfied. 
For diffusion-limited ion association, 
the linearization condition is rewritten as 
\begin{align}
 \frac{1}{4} \frac{(D_+ + D_-)^2}{D_+ D_-}
\frac{1}{1- \exp(-|r_c|/R)} >1. 
\label{eq:linearizationcon}
\end{align}
using Eq. (\ref{eq:barka_1}) and is always satisfied. 
The diffusion--limited association is commonly adopted for weak electrolytes 
because the ion association rapidly proceeds due to the Coulombic interaction 
without screening; \cite{Onsager_34,Onsager_38}
the screening by an ionic atmosphere only occurs in strong electrolytes. 

In calculating the diffusion--limited association rate, the concentration of counterions 
at the distance orders of magnitude larger than the Onsager radius is assumed to be homogeneous. \cite{Debye_42,Eigen_58,Onsager_34}
On the other hand, the density profiles obtained by solving the 1--dimensional e-PNP equations 
are characterized by the Debye length. 
If Onsager radius is orders of magnitude smaller than the Debye length, 
the combined approach could be utilized.  
The separation of length scales is satisfied 
when the relative dielectric constant is sufficiently high. 
For water, the Onsager length is $0.7$ nm which is orders of magnitude smaller than 
the Debye length given by $1$ $\mu$m. 

Regarding the dissociation rate constant $k_d$, 
we did not take into account the effect of electric fields on the ion dissociation rate. 
Strictly speaking, 
the dissociation rates of geminate ions have stronger electric field dependence compared to the association rates 
in weak electrolytes. \cite{Onsager_34,Traytak_91,Isoda_94,Hilczer_10}
However, the field dependence of the dissociation rates is negligibly small in the absence of external fields because 
the largest internal electrostatic potential difference 
in the spatial extent of the Debye length 
is on the same scale as thermal energy at room temperature.  
The linearization condition will be numerically justified in the next section.

%%%%%%%%%%%%%%%%%%%%%%%%%%%%%%%%%%%%%%
\section{Numerical justification of the linearization condition}
\label{sec:V}

In this section, we confirm the linearization condition numerically.  
In a steady state, 
we numerically solved Eqs. (\ref{eq:DLNp1})--(\ref{eq:BCC_xi})  
by the relaxation method. 
We present the numerical results obtained using the boundary condition including the effect of the electrode. 

As a concrete example, we studied the photo-induced electrochemical
reduction of water at electrode surfaces yielding hydrogen. 
Water is a weak electrolyte, and the dissociation of a small amount of water can be represented by 
the association and dissociation reaction terms in Eqs. (\ref{eq:sum1_0}) and (\ref{eq:change1_0}). 
The diffusion constants of H$_3$O$^+$ and OH$^-$ are 
$9.4 \times 10^{-9}$ m$^2$/s and $5.3 \times 10^{-9}$ m$^2$/s, respectively. \cite{Silbey}
By substituting 
$D_+=9.4 \times 10^{-9}$ m$^2$/s and $D_-=5.3 \times 10^{-9}$ m$^2$/s into Eq. (\ref{eq:DLD}), 
we obtained $\bar{D}=6.78 \times 10^{-9}$ m$^2$/s and $\Delta=0.28$. 
 Using $
n_b=1\times 10^{-7} \mbox{mol/L}
$, 
the Debye length was obtained as $\lambda_D=9.8 \times 10^{-7} \sim 10^{-6}$ m. 
The linearization condition of the e-PNP equations is given by $\bar{k}_a>1$, 
where the rate of ion association given by 
$k_a n_b$ is much higher than the inverse of the time required to diffuse over the Debye length given by 
$\bar{D}/\lambda_D^2$. 
In this section, we study the case when $\bar{k}_a=2.0$ and $\bar{k}_a=0.02$; 
the linearized results shown in Sec. \ref{sec:III} could be applicable to the case of $\bar{k}_a=2.0$ but not to the case of $\bar{k}_a=0.02$. 
For brevity, the case of $\bar{k}_a=2.0$ is classified as weak electrolytes and that of $\bar{k}_a=0.02$ 
is classified as strong electrolytes. 
In weak electrolytes, 
the intrinsic association rate constants are large, and the dissociation rate constants are small. 
The linearization condition is obtained when the association rate is diffusion limited 
because the intrinsic association rate constant is large. 
The linearization condition is most likely satisfied for weak electrolytes.  
On the contrary, the association rate constants of strong electrolytes are small and can be ignored. 

%%%%%%%%%%%%%%%%%%%%%%%%%%%%%%%%%%%%%%%
\begin{figure}
\centerline{
\includegraphics[width=0.6\columnwidth]{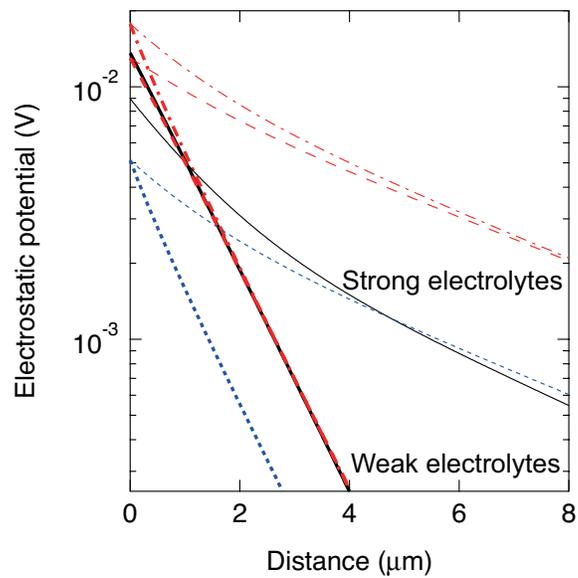}
}
\caption{(Color online) Electrostatic potential as a function of distance from 
the electrode when $N_s=0.5$. Thick lines indicate $\bar{k}_a=2.0$ (weak electrolytes) and 
thin lines indicate $\bar{k}_a=0.02$ (strong electrolytes). 
The solid lines represent the numerical solutions. 
(Red) dash-dotted lines represent the results of the Nernst equation given by Eq. (\ref{eq:Nernstprofile}).
(Red) long dashed lines represent the results of Eq. (\ref{eq:23}).
 (Blue) short dashed lines represent the results of the Henderson--Planck equation given by 
Eq. (\ref{eq:V_dcn}). 
}
\label{fig:NHP}
\end{figure}
%%%%%%%%%%%%%%%%%%%%%%%%%%%%%%%%%%%%%%%

In Fig. \ref{fig:NHP}, we compare the results of the Nernst equation, the Henderson--Planck equation, and Eq. (\ref{eq:23}). 
The electrostatic potential profile of the Nernst equation 
can be expressed as 
\begin{align}
V_{\rm N}(x)= \frac{k_{\rm B} T}{e} \ln\frac{n_b}{n_+(x)}, 
\label{eq:Nernstprofile}
\end{align}
which is equivalent to the local equilibrium assumption given by
\begin{align}
\frac{n_+(x)}{n_b}=\exp\left[-\frac{eV_{\rm N}(x)}{k_{\rm B} T} \right], 
\label{eq:dimN2}
\end{align}
where ion mass flows are assumed to be absent. 
Intuitively, 
the Nernst equation could be thought of as applicable to weak electrolytes 
when the ion discharge reaction at the electrode is slow so that the resultant deviation from charge neutrality 
near the surface is recovered by bulk reactions rather than ion diffusion. 
Indeed, we can show that the electrostatic potential profile given by Eq. (\ref{eq:23}) is obtained by 
linearizing Eq. (\ref{eq:Nernstprofile}) by assuming $[n_b-n_+(x)]/n_b \ll 1$ as follows: 
\begin{align}
V_{\rm N}(x)&= -\frac{k_{\rm B} T}{e} \ln
\left(1+\frac{n_+(x)-n_b}{n_b} \right) \\
&\approx \frac{k_{\rm B} T}{e} \left( 1- \frac{n_+(x)}{n_b}
\right) . 
\label{eq:Nernstprofilelinearized}
\end{align}
However, it should be reminded that Eq. (\ref{eq:Nernstprofilelinearized}) is obtained from the e-PNP equations when 
the linearization condition $\bar{k}_a>1$ is satisfied regardless of the condition on $[n_b-n_+(x)]/n_b$. 
In other words, if $\bar{k}_a>1$ is the linearization condition in the presence of diffusion, 
Eq. (\ref{eq:Nernstprofilelinearized}) should be satisfied even when $[n_b-n_+(x)]/n_b \sim 1$.

The Nernst equation is obtained by assuming that ion mass flows are absent.
The situation is irrelevant for strong electrolytes 
because electrode reactions induce ion mass flows in particular in the absence of ionization and association of ions. 
In strong electrolytes, ions are always dissociated and the terms that represent association and dissociation reactions on the 
right-hand sides of Eqs. (\ref{eq:NPR1}) and (\ref{eq:NPR2}) should be absent.
As mentioned above, linearization fails for strong electrolytes, where $\bar{k}_a < 1$. 
In the absence of association and dissociation reactions, 
approximate analytical solutions of the Poisson--Nernst--Planck equations have been studied more than decades. 
As shown in Appendix A and B, 
the potential difference 
known as the diffusion potential is obtained, \cite{Henderson_07,Bard_book_01,Dickinson2011,Jackson} 
\begin{align}
V_{\rm HP} (x)&= - \int_{x=L}^{x=0} dx E(x)=\frac{k_{\rm B} T}{e}
\frac{D_+ - D_- }{D_+ + D_- }
\ln \left(\frac{n_b}{n_+ (x)}\right).
\label{eq:V_dcn}
\end{align}
Equation (\ref{eq:V_dcn}) is also called the Henderson--Planck equation. 
In strong electrolytes, 
the potential difference is reduced from the Nernst equation 
by the factor given by $(D_+ - D_-)/(D_+ + D_-)$.
The Henderson--Planck equation represents the diffusion potential caused by 
different diffusion constants of anions and cations. \cite{Henderson_07,Bard_book_01,Dickinson2011,Jackson}

Coming back to Fig. \ref{fig:NHP}, we notice that 
the result of the Nernst equation is closer to the numerical result than that of the Henderson--Planck equation 
when $\bar{k}_a=2.0$ (weak electrolytes). 
However, the result of the Nernst equation deviates from the numerical result near the electrode. 
The potential profile obtained from Eq. (\ref{eq:23}) [or equivalently Eq. (\ref{eq:Nernstprofilelinearized})]
is close to the numerical result for all distances. 
Equation (\ref{eq:23}) was obtained by linearization of full transport equations and the Poisson equation. 
This result is valid under the linearization condition given by $\bar{k}_a>1$. 
The Nernst equation was obtained by assuming local equilibrium, where mass currents were not taken into account. 
The deviation found near the electrode could be caused by mass currents. 
When charge neutrality is broken near the electrode surface, 
both ionization and ion transport occur to recover charge neutrality. 
Because of ion transport, 
the local equilibrium assumption used to obtain the Nernst equation is violated, particularly 
near the electrode surface. 
The Henderson--Planck equation is derived for strong electrolytes, and 
its results cannot be applied to weak electrolytes. 

When $\bar{k}_a=0.02$,  the linearization condition $\bar{k}_a>1$ is not satisfied and 
the numerical results confirm that the results of Eq. (\ref{eq:23}) are not valid for this case. 
For strong electrolytes indicated by $\bar{k}_a=0.02$,  
the result of the Henderson--Planck equation is closer to the numerical result than that of the Nernst equation. 
The Henderson--Planck equation is derived for strong electrolytes, where ionization is not taken into account. 
In contrast, 
the Nernst equation is derived by assuming local equilibrium, which can be violated by mass currents. 
We should note that breakdown of charge neutrality can be recovered only by mass currents 
in strong electrolytes. 

For $\bar{k}_a=0.2$,  the numerical results deviate from both the results of Eq. (\ref{eq:23}) and the Henderson-Planck equation.
(Data not shown)
The results are consistent with the linearization condition given by $\bar{k}_a>1$.

%%%%%%%%%%%%%%%%%%%%%%%%%%%%%%%%%%%%%%%
\begin{figure}
\centerline{
\includegraphics[width=0.6\columnwidth]{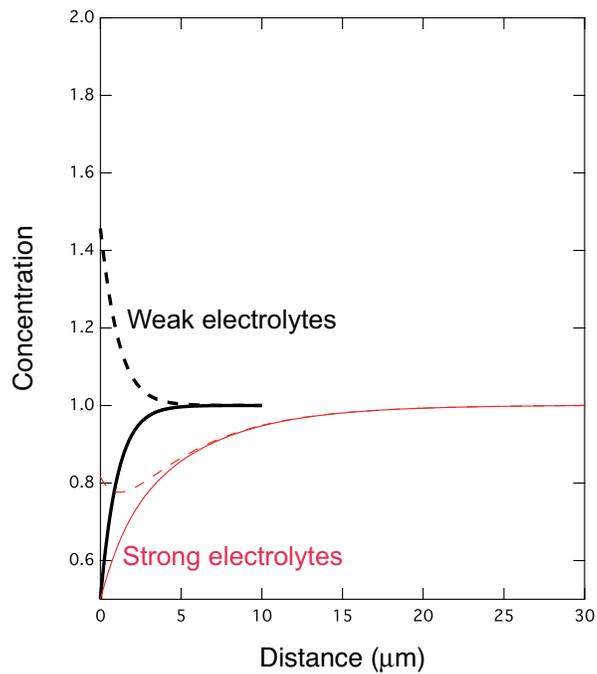}
}
\caption{(Color online) Concentrations of cations and anions as a function of distance from 
the electrode when $N_s=0.5$. Thick black lines indicate $\bar{k}_a=2.0$ (weak electrolytes) and 
thin (red) lines indicate $\bar{k}_a=0.02$ (strong electrolytes).
The solid lines represent cation concentrations and dashed lines represent anion concentrations. 
}
\label{fig:CA}
\end{figure}
%%%%%%%%%%%%%%%%%%%%%%%%%%%%%%%%%%%%%%%

In Fig. \ref{fig:CA}, we show concentrations of cations and anions as a function of distance from the electrode surface
 in strong and weak electrolytes. 
In weak electrolytes, both cation and anion concentrations deviate from the bulk concentration 
when the distance from the electrode is within a few Debye lengths. 
The result is consistent with the previous result obtained by assuming weak external electric field. \cite{AOKI_13}
Concentration profiles of cations are long-ranged in strong electrolytes. 
Interestingly, the cation and anion concentrations were different when 
the distance from the electrode was within a few Debye lengths in both strong and weak electrolytes. 
In other words, charge density caused by breakdown of charge neutrality was observed 
approximately within the surface layer to a depth of a few Debye lengths.
 The anion concentration profile is similar to that of cations outside this layer 
 in strong electrolytes due to charge neutrality.  
The difference between the spatial range of charge density and 
that of mass density for strong electrolytes has been pointed out. \cite{Mafe_86}
Although small, 
the increase of anion concentration near the electrode surface seen in Fig. \ref{fig:CA} 
could be caused by ionization of a few 
unionized molecules 
present in the strong electrolyte characterized by $\bar{k}_a=0.02$. 
We conclude that the linearization condition of the e-PNP equations is given by $\bar{k}_a>1$.

%%%%%%%%%%%%%%%%%%%%%%%%%%%%%%%%%%%%%%%%%%%%%%%%%%%%%%%%%%%%%%%%%%%%%%
% Both cation reduction and anion oxidation %%%%%%%%%%%%%%%%%%%%%%%%%%%%%%%%%%%%%%%%%%%%%%%%%%%%%%%%%%%%%%
%%%%%%%%%%%%%%%%%%%%%%%%%%%%%%%%%%%%%%%%%%%%%%%%%%%%%%%%%%%%%%%%%%%%%%
%\setcounter{equation}{0}
\section{Both cations and anions are discharged at the electrode surface}
\label{sec:BC} 
%\subsection{In the absence of electrolytic dissociation}
As indicated in Eq. (\ref{eq:deltaSsol}), 
the total charge density gradient could be scaled by 
the factor $1/\sqrt{2 \bar{k}_a}$ from the Debye length when  
the total charge density drops by allowing anions to be discharged as well as 
 cations at the electrode surface as shown in Fig. \ref{fig1} (b). 
In this section, we consider such situation.  

%%%%%%%%%%%%%%%%%%%%%%%%%%%%%%%%%%%%%%%
\begin{figure}
\centerline{
\includegraphics[width=0.6\columnwidth]{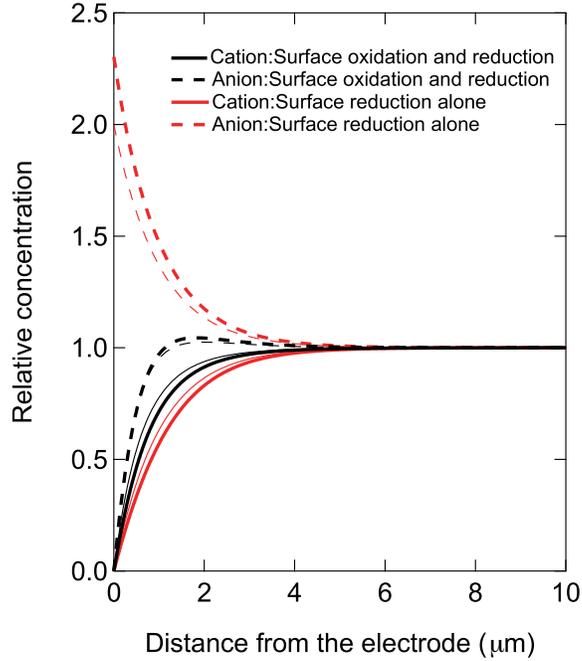}
}
\caption{Concentrations of cations and anions as a function of distance from 
the electrode. 
The solid lines indicate cations and the dashed lines indicate anions. 
The red lines represent the case when the perfectly reactive boundary condition 
($k_s \rightarrow \infty$ and $n_s=0$) is assumed for cations with inert anions. 
The black lines represent the case when anions are also perfectly reactive at the electrode surface.  
The thick lines represent the exact numerical results.  
The thin lines represent the approximate analytical results.  
The thin solid red line, dashed red line, solid black line and dashed black line is obtained from Eq. (\ref{eq:24}),  
Eq. (\ref{eq:24_1}), Eq. (\ref{eq:Npboth}), and Eq. (\ref{eq:Nmboth}), respectively.
}
\label{fig:c}
\end{figure}
%%%%%%%%%%%%%%%%%%%%%%%%%%%%%%%%%%%%%%%

The boundary conditions at $x=0$ ($y=0$) given by Eq. (\ref{eq:BCmm2}) changes into 
\begin{align}
n_- (0)&=n_{s-}.   
\label{eq:BCmm2b}
\end{align}
As in Eq. (\ref{eq:BCmp1})
$n_{s-}$ represents the net effect of surface reactions on $n_- (0)$.    

The boundary condition at $y=0$ obtained from Eq. (\ref{eq:BCmm2b}) can be written as 
\begin{align}
\left. \delta S+\delta C +2(1-N_{s-})\right|_{y=0}=0,
\label{eq:BC_oxidation}
\end{align}
where $N_{s-}=n_{s-}/n_b$ as $N_s$ in Eq. (\ref{eq:DLN}). 
Similarly, we have $\left. \delta S+\delta C +2(1-N_s)\right|_{y=0}=0$ from 
Eq. (\ref{eq:BCSm1_0o}).
By substituting Eqs. (\ref{eq:deltaSsol})-(\ref{eq:deltaCsol}) into the above boundary conditions, we 
find $C_s=-2+N_s+N_{s-}$ and $C_c=N_s-N_{s-}-(2-N_s-N_{s-})2 \bar{k}_a \Delta/(2\bar{k}_a-1)$. 
For the ideal case of $N_s=N_{s-}=0$, we obtain, 
\begin{align}
N_+&= 1- \left(1-\frac{2 \bar{k}_a \Delta}{2\bar{k}_a-1} \right) \exp \left(-\sqrt{2 \bar{k}_a}\, y\right) -
\frac{2 \bar{k}_a \Delta}{2\bar{k}_a-1} \exp \left(-y\right),
\label{eq:Npboth}\\
N_-&= 1- \left(1+\frac{2 \bar{k}_a \Delta}{2\bar{k}_a-1} \right) \exp \left(-\sqrt{2 \bar{k}_a}\, y\right) +
\frac{2 \bar{k}_a \Delta}{2\bar{k}_a-1} \exp \left(-y\right),
\label{eq:Nmboth}\\
\xi&= -\frac{\sqrt{2 \bar{k}_a}\, \Delta}{2\bar{k}_a-1}\exp \left(-\sqrt{2 \bar{k}_a}y\right) +
\frac{2 \bar{k}_a \Delta}{2\bar{k}_a-1} \exp \left(-y\right), 
\label{eq:xiboth}\\
V&= \frac{k_{\rm B} T}{e} \frac{\Delta}{2\bar{k}_a-1}
\left[ 2 \bar{k}_a\exp \left(-y\right)- \exp \left(-\sqrt{2 \bar{k}_a}\, y\right) 
\right].
\label{eq:Vboth}
\end{align}
The factor $\sqrt{2 \bar{k}_a}$ indicates 
the length scale given by the inverse of $\sqrt{2 k_a n_b/\bar{D}}$.  
For diffusion limited ion association in weak electrolytes, 
$\bar{k}_a>1$ holds and we can estimate 
that $\sqrt{\bar{D}/(2k_a n_b)}$ might be smaller than the Debye length 
because $\bar{k}_a>1$ can be rewritten as $\lambda_D>\sqrt{\bar{D}/(k_a n_b)}$. 
The concentration gradients are essentially characterized by the length scale given by $\sqrt{\bar{D}/(2k_a n_b)}$ 
rather than the Debye length when both cations and anions are discharged at the electrode. 
Interestingly, the electrostatic potential difference is essentially characterized by 
the Debye length because $ 2\bar{k}_a>1$ holds in Eq. (\ref{eq:Vboth}). 
The result is reasonable because the electrostatic potential difference originates from 
the capacitance effect caused by the charge density near the electrode; 
the charge density gradients are 
characterized by the Debye length. 

As shown above, the length scale given by $\sqrt{\bar{D}/(2 k_a n_b)}$ characterizes the 
ion concentration gradients. 
In an entirely different context but using the similar method, it was found that density profiles could be characterized 
by the association reaction-diffusion length scale in certain multi-ionic solutions. \cite{Brown_17}
$\sqrt{\bar{D}/(2 k_a n_b)}$ represents the length scale of the diffusive migration within the life-time of ions 
which will disappear as a consequence of the ion association reaction.
The length scale is called the Kuramoto length. \cite{Kuramoto_73,Kuramoto_74,Nitzan_74,Nicolis_84,Kitahara_90}  
The Kuramoto length characterizes the spatial correlation of density fluctuations around a uniform concentration state. 
The density fluctuations beyond the length scale given by the Kuramoto length are independent. 
Here, we show that the ion density drop at the electrode surface can be localized within the Kuramoto length.
The Kuramoto length can be expressed using the diffusion-controlled rate shown in Sec. \ref{sec:IV} as 
\begin{align}
\sqrt{
\frac{D_+ D_-
[1- \exp(-|r_c|/R)]}{4\pi (D_++D_-)^2 |r_c| n_b}
}. 
\label{eq:Kuramoto:diff}
\end{align}
In water, the Kuramoto length is given by $\lambda_D/2\approx 0.5$ \mbox{$\mu$m}.

\begin{figure}
\centerline{
\includegraphics[width=0.6\columnwidth]{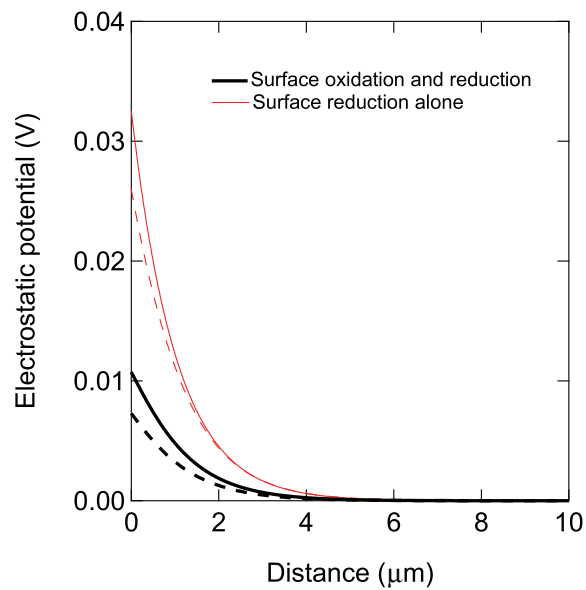}
}
\caption{(Color online) Electrostatic potential as a function of distance from 
the electrode. 
The thin (red) lines represent the case when the perfectly reactive boundary condition 
($n_s=0$) is assumed for cations with inert anions. 
The thick (black) lines represent the case when anions are also perfectly reactive at the electrode surface.  
The solid lines indicate the exact numerical results. 
The dashed lines represent the approximate analytical results.  
The (red) dashed line represents $V(y)= (k_{\rm B} T/e) (1-N_+)$ obtained from Eq. (\ref{eq:23}) using 
the numerical results for cation concentration. 
The (black) dashed line is obtained from  Eq. (\ref{eq:Vboth}).  
}
\label{fig:ELV}
\end{figure}

We compare the above results with the numerical solutions as in the previous section. 
As an example, we consider photo-induced water splitting reactions by photocatalysts. 
We set $\bar{k}_a=2.0$ and the other parameters are the same as those in the previous section. 
In Fig. \ref{fig:ELV}, we show the cation and anion concentration profiles as a function of distance from 
the electrode. 
We can see that the above analytical solutions satisfactory reproduce the numerical results.
Note that the concentration gradients are short ranged when both cations and anions are reactive 
at the electrode surface compared to the case when only cations are reactive. 
The difference reflects the fact that the concentration gradients are characterized by the Kuramoto length for the former and 
the Debye length for the latter.  
The steeper concentration gradients generated the larger ion currents by diffusion. 
In this respect, as long as side effects such as back reactions can be ignored, 
the photo-catalysts that are reactive to both cations and anions are desirable.

%%%%%%%%%%%%%%%%%%%%%%%%%%%%%%%%%%%%%%%
%%%%%%%%%%%%%%%%%%%%%%%%%%%%%%%%%%%%%%%

In Fig. \ref{fig:ELV}, we show electrostatic potential as a function of distance from 
the electrode. 
Again, the analytical expressions reproduce the numerical results. 
When only cations are reduced, and anions are not reactive at the electrode, 
the electrode potential is higher than the electrostatic potential in the bulk water, and 
the result can be interpreted as overpotential. 
When both cations and anions are reactive at the electrode, 
the difference in the electrode potential is reduced and 
is proportional to the difference between the cation diffusion constant and the anion diffusion constant 
as shown in Eq. (\ref{eq:Vboth}). 
Depending on the difference between the diffusion constants, the electrostatic potential difference can be negative. 
For any case, the electrostatic potential gradient is characterized by the same length scale given by 
the Debye length. 
The situation is very different from the ion concentration profiles, where the length scale is characterized by 
the Kuramoto length when both cations and anions are discharged.

%%%%%%%%%%%%%%%%%%%%%%%%%%%%%%%%%%%%%%%%%%%%%%%%%%%%%%%%%%%%%%%%%%%%%%
% Conclusion %%%%%%%%%%%%%%%%%%%%%%%%%%%%%%%%%%%%%%%%%%%%%%%%%%%%%%%%%%%%%%
%%%%%%%%%%%%%%%%%%%%%%%%%%%%%%%%%%%%%%%%%%%%%%%%%%%%%%%%%%%%%%%%%%%%%%
%\setcounter{equation}{0}
\section{Conclusion}
\label{sec:Conclusion} 

The results of this paper are related to photo-electrochemical (PEC) water splitting. 
Although theoretically estimated PEC conversion efficiency exceeds 10\%, \cite{Bolton86,Hanna06,Weber84,Weber86,Seitz14,Rocheleau97} 
such high efficiency has not yet been achieved without an external energy supply. \cite{McCrory15,Bolton86,Cox14} 
Unlike photovoltaics, 
the PEC conversion efficiency can be limited by electrochemical processes occurring at the interface between an electrode and 
electrolyte even if a semiconductor with a suitable band gap is used. 
There could be two main limiting  factors for PEC conversion efficiency related to the electrode--electrolyte interface. 

The first limitation 
of PEC conversion efficiency could be that ion mobilities can be lower  than electron mobilities in 
semiconductors. 
Steep ion concentration gradients would be preferable to induce ion currents 
to resolve the inhomogeneous ion distribution. 
We have shown that the concentration gradients are characterized by the Debye length 
when cations are reduced at the electrode, and the anions are inert for 
1:1 monovalent weak electrolytes.  
The concentration gradient can be approximately given by 
$n_b/\lambda_D$. 
By using $D_+\approx 10^{-8}$ m$^2$/s, $\lambda_D \approx 10^{-6}$ m, 
the ion current density can be expressed as 
\begin{align}
e D_+ \frac{n_b}{\lambda_D} \times 10^5 \mbox{ A/cm$^2$} 
=0.01 \mbox{ mA/cm$^2$}, 
\label{eq:ioniccurrent}
\end{align}
when the boundary condition is given by 
$n_+ (0)=0$ corresponding to the case where the back reaction can be ignored because of fast forward reactions 
at the interface, 
where $96500$ C/mol is approximated by $10^5$ C/mol. 
The above result matches with that of linearization given by Eq. (\ref{eq:current_liear_0}) and 
is 30\% smaller than the exact numerical result. 
We have also shown that the concentration gradient is characterized by 
the Kuramoto length 
when both cation reduction and anion oxidation reactions occur at the electrode 
for 1:1 monovalent weak electrolytes.  
The Kuramoto length is about the half of the Debye length. 
As a result,   
the ion current can be twice larger than that obtained when anions are not oxidized at the electrode surface. 
However, 
to realize 10\% PEC conversion efficiency under 1 sun, a current density of $10$ mA cm$^{-2}$ is required. 
 This result indicates that ion currents may limit the overall efficiency because of the small diffusion constant of H$_3$O$^+$ 
compared with that of electrons. 
For photo-induced splitting of pure water, 
recent experimental setups have used 
internal convection caused by placing cathodic and anodic electrodes in parallel configurations. \cite{Wang_16,Wang_17}
In this configuration, an extra energy source to generate convection was not needed. 
In principle, convective flows can be considered by introducing the hydrodynamic equation into 
the Nernst--Planck equations. \cite{Kontturi_08}
An extension of the present work to include convective flows will be important to study 
the case when the current density is high. 

The second limitation could be overpotential related to a charge density gradient near the electrode.  
Our results indicate that the overpotential is at most on the same scale as thermal energy at room temperature 
and can be further reduced  
when both cation reduction and anion oxidation reactions occur at the electrode 
for 1:1 monovalent weak electrolytes.  
The overpotential is caused by a charge density gradient whose spatial extent is 
characterized by the Debye length. 
It should be noted that 
the charge density gradient can be estimated from the pH gradient 
only for pure water. 
In general,  
charge density cannot be estimated from pH alone in the presence of other ion species.
  
For simplicity, we considered the case that 
molecules could be ionized to monovalent ions 
in the absence of buffer electrolytes composed of weak acid and/or weak base. 
We also did not consider supporting electrolytes to increase the conductivity of the bulk solution. 
For photo-induced water-splitting reactions,  multi-ionic solutions are commonly used to 
control electrostatic potentials and engineer the pH of electrolytes. \cite{Vermaas_15,Xiang_16,Jin_14}
There, large-scale pH gradients were observed. 
The observed large-scale pH gradients 
could be related to the length-scale separation 
between charge density profiles and concentration profiles, which could be possible for multi-ionic solutions. 
The effect of multiple ionic species on ion currents could be taken into account 
using new numerical methods developed to  
calculate junction potentials,\cite{Dickinson2011,Urtenov_07} where the charge neutrality condition is unnecessary. 
The analytical method applied in this paper should be extended to multi-ionic solutions in the future.    

%%%
\section*{Acknowledgement}
This work was supported by the ``Research Project for Future Development: Artificial
Photosynthetic Chemical Process (ARPChem)'' (METI/NEDO, Japan: 2012-2022).

\appendix
%%%%%%%%%%%%%%%%%%%%%%%%%%%%%%%%%%%%%%%%%%%%%%%%%%%%%%%
%\newpage
%\renewcommand{\theequation}{A.\arabic{equation}}  
%\setcounter{equation}{0}  % reset counter     
%\section*{Appendix A. 
\section{Derivation of Eq. (\ref{eq:V_dcn})}
We performed a systematic expansion in $1/\tau$ when $\bar{k}_a=0$ by applying the approach developed to calculate 
liquid junction potential to our problem of surface reactions. \cite{Jackson}
When concentration change is restored only by diffusion, a relevant variable is 
$z= y/\sqrt{\tau}$. 
In terms of the new variable, 
Eq. (\ref{eq:DLNp1})--(\ref{eq:DLNE1}) can be rewritten using 
$\delta S = N_+ + N_- -2$, $\delta C= N_+ - N_-$, and $z$ 
by applying the chain rule for partials as 
\begin{align}
\frac{1}{\tau} \frac{\partial^2 \delta S}{\partial z^2} - 
\frac{\partial}{\partial z} \left( \frac{\xi }{\sqrt{\tau}}\delta C\right) &= 
\frac{\partial}{\partial \tau} \left( \delta S - \Delta \delta C \right) 
-\frac{z}{2\tau} \frac{\partial }{\partial z}\left( \delta S - \Delta \delta C \right)
,
\label{eq:sum2_0}\\
\frac{1}{\tau} \frac{\partial^2 \delta C}{\partial z^2} - 
\frac{\partial}{\partial z} \left[ \frac{\xi }{\sqrt{\tau}} (\delta S+2) \right] &= 
\frac{\partial}{\partial \tau} \left( \delta C - \Delta \delta S \right) 
-\frac{z}{2\tau}\frac{\partial }{\partial z}\left( \delta C - \Delta \delta S \right)
,
\label{eq:change2_0}\\
\frac{\partial}{\partial z} \frac{\xi }{\sqrt{\tau}} &= \frac{1}{2} \delta C.
\label{eq:xiC2_0}
\end{align}

The diffusion may give rise to algebraic $\tau$ dependence, 
\begin{align}
\delta S(z,\tau)&= \delta S(z)^{(0)}+\delta S(z)^{(1)}/\tau+\delta S(z)^{(2)}/\tau^2 +\cdots,
\label{eq:deltaStau_0}\\
\delta C(z,\tau)&= \delta C(z)^{(0)}+\delta C(z)^{(1)}/\tau+\delta C(z)^{(2)}/\tau^2 +\cdots,
\label{eq:deltaCtau_0}\\
\frac{\xi(z,\tau)}{\sqrt{\tau}} &= \frac{\eta^{(1)}(z)}{\tau}+\frac{\eta^{(2)}(z)}{\tau^2}+\cdots. 
\label{eq:eta_0}
\end{align}
$\eta^{(0)}(z)$ should be zero; otherwise, $\xi(z) \sim \sqrt{\tau} \eta^{(0)}(z)$ diverges as 
$\tau \rightarrow \infty$. 

Now, we study the lowest-order solution given by the order of $1/\tau^0$. 
From Eq. (\ref{eq:xiC2_0}), the charge neutrality condition is satisfied; i.e., 
\begin{align}
\delta C(z)^{(0)}=0. 
\label{eq:cn1}
\end{align}

By substituting Eqs. (\ref{eq:deltaStau_0}) and (\ref{eq:deltaCtau_0}) 
into Eqs. (\ref{eq:sum2_0})-(\ref{eq:xiC2_0}), 
we obtained $S^{(0)}(z) $ and $\eta^{(1)}(z) $ of the order of $1/\tau$ as
\begin{align}
\frac{\partial^2 \delta S^{(0)}}{\partial z^2}  &= 
-\frac{z}{2}\frac{\partial }{\partial z} \left( \delta S^{(0)}  \right),
\label{eq:sum3_0}\\
- 
\frac{\partial}{\partial z} \left[ \eta^{(1)}(z) (\delta S^{(0)}+2) \right] &= 
 \Delta\frac{z}{2}\frac{\partial }{\partial z} \delta S^{(0)}.
 \label{eq:change3_0}
\end{align}
Boundary conditions at $z=L/(\lambda_D\sqrt{\tau})$ are given by
\begin{align}
\left. \delta S^{(0)}\right|_{z=L/(\lambda_D\sqrt{\tau})}=0, 
\label{eq:BCS1_0}\\
\left. \delta C^{(0)}\right|_{z=L/(\lambda_D\sqrt{\tau})}=0. 
\label{eq:BCC2_0}
\end{align}

The boundary condition at $y=0$ is expressed as
\begin{align}
\left.\delta S^{(0)}\right|_{z=0}\approx2\left(N_s-1\right). 
\label{eq:BCCsimple}
\end{align}
The potential difference can be obtained from  
\begin{align}
V(z)=  E_0 \lambda_D \int_{z}^{\infty} dz_1 \eta^{(1)} (z_1), 
\label{eq:V1_0}
\end{align}
and the total potential difference between the electrode surface and bulk 
is  $V= V (0)$.

First, we integrated Eq. (\ref{eq:sum3_0})  and obtained 
\begin{align}
\frac{\partial }{\partial z} \left( \delta S^{(0)}  \right)
= A_1\exp\left(- \frac{z^2}{4} \right), 
\label{eq:sold2_0}
\end{align}
where $A_1$ is an integration constant.  
By integrating the above equation, we have 
\begin{align}
\delta S^{(0)}(z) = A_0 + A_1 \int_0^z d z_1  \exp\left(- \frac{z_1^2}{4} \right), 
\label{eq:sold3_0}
\end{align}
where $A_0$ is an integration constant.

We note that 
$\int_0^\infty d z_1  \exp\left(- \frac{z_1^2}{4} \right)= \sqrt{\pi}$. 
Using the boundary conditions given by Eqs. (\ref{eq:BCS1_0}) and (\ref{eq:BCCsimple}), we obtained
\begin{align}
A_0 &=2\left( N_s- 1\right),   
\label{eq:sold5_0}\\
A_1&=2\left( 1-N_s\right)/\sqrt{\pi}. 
\label{eq:sold6_0}
\end{align}
By substituting these equations into Eq. (\ref{eq:sold3_0}), it became 
\begin{align}
\delta S^{(0)}(z) = 2\left(1-N_s\right)\left[-1 + \mbox{erf}(z/2)\right], 
\label{eq:sold7_0}
\end{align}
where we used 
$\mbox{erf}(z/2)=
\left(1/\sqrt{\pi}\right)\int_0^z d z_1  \exp\left(- z_1^2/4 \right)$. \cite{Abramowitz}
Note that $\lim_{z \rightarrow \infty}  \mbox{erf}(z/2)=1$ and 
we have $\lim_{z \rightarrow \infty} \delta S^{(0)}(z)=0$, as expected.

Finally, using Eq. (\ref{eq:sold2_0}), we obtained from Eq. (\ref{eq:change3_0})   
\begin{align}
\frac{\partial}{\partial z} \left[ \eta^{(1)}(z) (\delta S^{(0)}(z) +2) \right] 
&= 2\Delta\frac{1-N_s}{\sqrt{\pi}} \frac{\partial}{\partial z} \exp\left(- \frac{z^2}{4} \right)
.
\label{eq:sold9_0}
\end{align}
By further integration, we find 
\begin{align}
\eta^{(1)}(z) 
&= \frac{\Delta\frac{1-N_s}{\sqrt{\pi}} \exp\left(- \frac{z^2}{4} \right)+A_3}{
N_s + \left(1-N_s\right)\mbox{erf}(z/2) },  
\label{eq:sold11_0}
\end{align}
where we used
$\lim_{z \rightarrow \infty} \delta S^{(0)}=0$ and $A_3$ is an integration constant. 

We consider the case that the electric field at the infinite distance from the electrode is zero.  
The $L \rightarrow \infty$ limit of $z$ is given by $z \rightarrow \infty$ 
because we have $z=y/\sqrt{\tau}$, $y=x/\lambda_D$, and $x=L$. 
Noting that 
the dimensionless field is denoted by $\eta^{(1)}$, 
we obtained $A_3=0$ from Eqs. (\ref{eq:sold11_0}) and (\ref{eq:BCEM}). 

By substituting Eq. (\ref{eq:sold11_0}) into Eq. (\ref{eq:V1_0}), the voltage difference was calculated as 
\begin{align}
V(z)&= E_0 \lambda_D \Delta\int_{z}^{\infty} dz \frac{\frac{1-N_s}{\sqrt{\pi}} \exp\left(- \frac{z^2}{4} \right)}{
N_s + \left(1-N_s\right)\mbox{erf}(z/2)}\\
&=
\left. 
 E_0 \lambda_D \Delta\ln \left[N_s + \frac{1-N_s}{\sqrt{\pi}} \int_0^{z'} d z_1  \exp\left(- \frac{z_1^2}{4} \right)
\right]
\right|_{z'=z}^{z'=\infty} 
\\
&=-E_0 \lambda_D \Delta\ln \left[N_s+\left(1-N_s\right) \mbox{erf}\left(\frac{z}{2}\right)
\right]. 
\label{eq:V2_0_o}
\end{align}
The total voltage difference between the electrode surface and bulk was obtained 
from $V(0)$ as 
\begin{align}
V_{\rm HP}&=-\frac{k_{\rm B} T}{e} \frac{D_+ - D_-}{D_+ + D_-}
\ln \left(N_s\right). 
\label{eq:V2_0}
\end{align}
Although the concentration of cations should be lower than that of anions near the electrode 
because of the reduction of H$_3$O$^+$ at the electrode surface, 
the charge neutrality condition expressed by 
Eq. (\ref{eq:cn1}) was obtained as 
the lowest-order solution from the expansion of $1/\tau$.

Conventionally, Eq. (\ref{eq:V2_0}) has been derived using the difference in chemical potentials at the boundaries 
of liquid junctions. \cite{Jackson}
 In the derivation in Appendix A, we used different boundary conditions, and 
the result indicates that the voltage change caused by electrolytic reactions at an electrode surface 
can be approximated by the Henderson--Planck equation in strong electrolytes. 

%\newpage
%\renewcommand{\theequation}{B.\arabic{equation}}  
%\setcounter{equation}{0}  % reset counter     
%\section*{Appendix B. 
\section{The electrostatic potential under charge neutrality condition}
Interestingly, 
the diffusion potential can be derived by assuming 
charge neutrality 
regardless of presence of association and dissociation reactions.
When the charges do not accumulate at the interface between the electrode and electrolyte, 
the positive and negative currents should be equal. 
The equality of the positive and negative currents in the steady state can be expressed as 
\begin{align}
D_+ \left[ \frac{\partial n_+}{\partial x}  - \left( 
\frac{eEn_+}{k_{\rm B} T} 
\right)
\right]
=
D_- \left[ 
\frac{\partial n_-}{\partial x}  + \left( 
\frac{eEn_-}{k_{\rm B} T} 
\right)
\right],
\label{eq:jbalance}
\end{align}
and the charge neutrality condition is given by $n_+=n_-$.
By introducing the charge neutrality condition, we obtain from Eq. (\ref{eq:jbalance}) \cite{Jackson}
\begin{align}
E&=\frac{k_{\rm B} T}{e}
\frac{D_+ - D_- }{D_+ + D_- } \frac{1}{n_+}\frac{\partial n_+}{\partial x}
, 
\label{eq:cn3}
\end{align}
and Eq. (\ref{eq:V_dcn}) 
known as the Henderson--Planck equation for 1:1 monovalent electrolytes even in the presence of association and dissociation reactions. 
The fact that 
Eq. (\ref{eq:23}) [or equivalently Eq. (\ref{eq:Nernstprofilelinearized})] is not obtained by assuming local charge neutrality 
indicates the subtle aspects of charge neutrality to calculate the electrostatic potential. 

%%%

\end{document}